\documentstyle[aps,twocolumn,epsf,floats]{revtex}

\newcommand{\lsi}{\raise0.3ex\hbox{$<$\kern-0.75em\raise-1.1ex\hbox{$\sim$}}}
\newcommand{\gsi}{\raise0.3ex\hbox{$>$\kern-0.75em\raise-1.1ex\hbox{$\sim$}}}
\newcommand{\lsim}{\mathop{\lsi}}
\newcommand{\gsim}{\mathop{\gsi}}
\renewcommand{\vec}[1]{{\bf #1}}
\newcommand{\stern}{{}^\ast\!}

\newcommand{\intv}[1]{\int\!\frac{d\Omega_{{\rm v}_{#1}}}{4\pi}}

\newcommand{\mref}[1]{(\ref{#1})}

\newcommand{\eq}[1]{Eq.\ \mref{#1}}
\newcommand{\fig}[1]{Fig.\ \ref{#1}}
\newcommand{\mlabel}[1]{\label{#1}}

\renewcommand{\(}{\left(}
\renewcommand{\)}{\right)}
\renewcommand{\[}{\left[}
\renewcommand{\]}{\right]}

\newcommand{\mmdebye}{m^2_{\rm D}}
\newcommand{\mdebye}{m_{\rm D}}
\newcommand{\tr}{{\rm t}}
\newcommand{\trace}{{\rm tr}}

\newcommand{\vA}{\vec{A}}
\newcommand{\vB}{\vec{B}}
\newcommand{\vE}{\vec{E}}

\newcommand{\ve}{\vec{e}}
\newcommand{\vk}{\vec{k}}
\newcommand{\vp}{\vec{p}}

\newcommand{\vv}{\vec{v}}
\newcommand{\vx}{\vec{x}}
\newcommand{\mal}{\!\cdot\!}

\newcommand{\deltav}{\delta^{(S^2)}}
\newcommand{\lav}{\langle\!\langle}
\newcommand{\rav}{\rangle\!\rangle}
\newcommand{\nn}{\nonumber}

\begin{document}
\twocolumn[\hsize\textwidth\columnwidth\hsize\csname
@twocolumnfalse\endcsname

\title{An effective theory for hot non-Abelian dynamics$^\dagger$}

\author{%
\hfill 
Dietrich B\"odeker$^*$
\hfill\raisebox{21mm}[0mm][0mm]{\makebox[0mm][r]{HD-THEP-98-35}}%
\raisebox{17mm}[0mm][0mm]{\makebox[0mm][r]{NBI-HE-98-29}}
}
\address{Institut f\"ur Theoretische Physik, Universit\"at Heidelberg, 
Philosophenweg 16, 
D-69120~Heidelberg, Germany
}

\date{\today}

\maketitle

\begin{abstract}
I try to explain some recent progress in understanding the
non-perturbative dynamics of hot non-Abelian gauge theories.  The
non-perturbative physics is due to soft spatial momenta $|\vec{p}|\sim g^2 T$
where $g$ is the gauge coupling and $T$ is the temperature. An
effective theory for the soft field modes is obtained
by integrating out the field modes with momenta of order $T$ and of
order $g T$ in a leading logarithmic approximation.  In this effective
theory the time evolution of the soft fields is determined by a local
Langevin-type equation.  This effective theory determines the
parametric form of the rate for hot electroweak baryon number
violation as $\Gamma = \kappa g^{10} \log(1/g)  T^4$. The
non-perturbative coefficient $\kappa$ is independent of the gauge
coupling and it can be computed by solving the effective equations of
motion on a lattice.

\end{abstract}

\vskip1.5pc]


\section{Introduction}

Let me start by formulating the problem I am going to discuss: How can
one calculate thermal expectation values like
\begin{eqnarray}
        C(t_1 - t_2) = \langle {\cal O}(t_1) {\cal O}(t_2)\rangle
        \mlabel{c}
\end{eqnarray}
in a non-Abelian gauge theory, when the leading order contribution is
due to spatial momenta of order $g^2 T$? The operator ${\cal
O}(t)$ is a gauge invariant function of the gauge fields
$A_\mu(t,\vx)$ at time $t$. When I said the leading order contribution
is due to momenta of order $g^2 T$, I referred to the Fourier
components of the gauge fields entering ${\cal O}(t)$.
Finally,
\begin{eqnarray}
        \langle \cdots \rangle = 
        \frac{\trace\{(\cdots)e^{- H/T}\}}{\trace\{e^{-H/T}\}}
\end{eqnarray}
denotes the thermal average, where $H$ is the Hamiltonian. 
As one may anticipate, it is not possible to compute such a correlation
function in perturbation theory. 

\footnotetext{$^\dagger$Talk presented 
at the 5th International Workshop on Thermal Field Theories 
and their Applications, Regensburg, Germany, August 1998}

\footnotetext{$^*$Address after October 1: The Niels Bohr Institute,
Blegdamsvej 17, DK-2100 Copenhagen \O, Denmark; e-mail: bodeker@nbi.dk}
\setcounter{footnote}{2}

This problem  arises in the context of electroweak baryon number
violation at temperatures, well above 100GeV, when the electroweak
symmetry is unbroken. Then the SU(2) gauge theory is very similar to hot
QCD. Fortunately, it is a bit simpler because the gauge coupling is
small.

In my talk I will try to explain how such correlation functions can be
computed at leading order in the gauge coupling \cite{letter}.  We
will see that this requires the use of an effective theory which
results from integrating out field modes with hard ($p\sim T$) and
semi-hard ($p\sim gT$) spatial momenta \footnote{For spatial vectors
I use the notation $k=|\vk|$. Four-vectors are denoted by $K^\mu =
(k^0,\vk)$ and I use the metric $K^2 = k_0^2 - k^2$.}. Due to the
smallness of the gauge coupling this can be done in perturbation
theory. This effective theory turns out to have a relatively simple
structure and it can be easily used for numerical calculations on a
lattice.

Most of the discussion will be restricted to a pure gauge theory, i.e.,
without matter fields. I will comment on the role of matter fields
when necessary.

\section{Hot electroweak baryon number violation}\mlabel{sec:baryon}

One of the most intriguing aspects of the electroweak theory is its
non-trivial vacuum structure which is related to baryon number violating
processes. These processes play an important role in understanding the
observed baryon-asymmetry of the universe \cite{rubakov}. 

Baryon number is not conserved in the electroweak theory due to the
chiral anomaly. The baryon number current satisfies \footnote{Baryon
plus lepton number is violated in the electroweak theory while baryon
minus lepton number is conserved.}
\begin{eqnarray}
        \partial_\mu j^\mu_B = n_f \frac{ g^2}{32\pi^2} \trace\(F\tilde{F}\)
        \mlabel{anomaly}
\end{eqnarray}
where $F$ is the SU(2) field strength tensor. The rhs of \eq{anomaly}
is a total derivative,
\begin{eqnarray}
        \trace\(F\tilde{F}\) = \partial_\mu K^\mu.
\end{eqnarray}
Thus, integrating \eq{anomaly} over 3-space and over time from $t_i$
to $t_f$ we can write the change of baryon number $B$ as
\begin{eqnarray}
        B(t_f) - B(t_i) = n_f ( N_{\rm CS}(t_f) - N_{\rm CS}(t_i))
        \mlabel{deltaB}
\end{eqnarray}
where 
\begin{eqnarray}
   N_{\rm CS} = \frac {g^2}{32\pi^2} \epsilon_{ijk} \int d^3x
  \left( F^a_{ij} A^a_k - \frac{g}{3} \epsilon^{abc} 
  A^a_i A^b_j A^c_k \right) 
  \mlabel{chern}
\end{eqnarray}
is the so called Chern-Simons winding number of the gauge fields.
Unlike $N_{\rm CS}$ itself, the difference $\Delta N_{\rm CS}=N_{\rm
  CS}(t_f) - N_{\rm CS}(t_i)$ is gauge invariant since it can be
written as a space-time integral of $F\tilde{F}$.

The Chern-Simons number is intimately related to the non-trivial
topology of the gauge-Higgs field configuration space.  There are paths in
this space leading from the vacuum to the vacuum which cannot
be continuously deformed to a single point. The minimal field energy
along such a path cannot be made arbitrarily small by continuous
deformation. There is a minimal energy barrier which has to be
crossed. The corresponding field configuration is the so called
sphaleron.  Following such a non-contractible path, the Chern-Simons
number changes by an integer.

The simplest mechanical analogue of this situation I can think of is
a rigid pendulum in a gravitational field which can swing all the way
around in a circle. The Chern-Simons number corresponds to
$\phi/(2\pi)$ where $\phi$ is the rotation angle.  The vacuum
corresponds to the pendulum at rest and a non-contractible loop
corresponds to net rotation of the pendulum by an integer times
$2\pi$.  

\begin{figure}[t]

\vspace*{-1.5cm}
 
\hspace*{1cm}
\epsfysize=20cm
\centerline{\epsffile{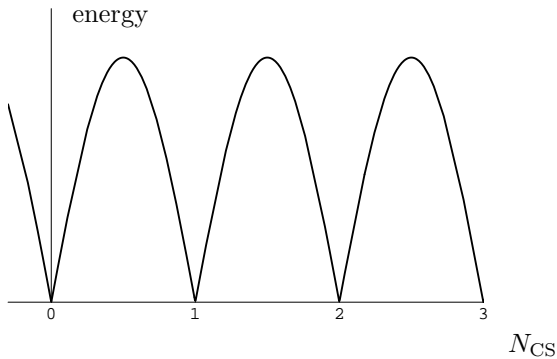}}

\vspace*{-14cm}
 
 \begin{picture}(50,50)(40,0)
\put(85,165){energy}
\put(250,40){$N_{\rm CS}$}
\end{picture}
\vspace*{-1cm}
\caption{One dimensional slice of the gauge-Higgs field configuration space.
  The vertical axis is the energy of a field configuration which has
  the minimal energy for given Chern-Simons winding number.}
\mlabel{fig:chern}
\end{figure}

The gauge-Higgs field configuration space is of course much more
complicated, it is infinite dimensional rather than one dimensional as
in the case of the pendulum.  \fig{fig:chern} displays a one
dimensional slice of this space. It
corresponds to a path on which the field configuration for a given
value of $N_{\rm CS} $ has a minimal energy.

If there were only gauge and Higgs fields, no\-thing
spectacular would happen in a vacuum to vacuum transition which
changes $N_{\rm CS} $.  However, in the presence of fermions, the
change of $N_{\rm CS} $ is accompanied by a change of baryon number
due to \eq{deltaB}.  

At zero temperature, $N_{\rm CS} $-changing
transitions are possible only by tunnelling. The
tunnelling probability is so small that it has no measurable physical
effect. 

This situation changes at high temperature. Then it becomes possible
to make such transitions by thermal activation across the energy barrier. The
suppression factor is then given by a Boltzmann factor rather than the
tunnelling amplitude.  For $T$ of the order of the electroweak phase
transition or crossover temperature $T_{\rm c}\sim 100$GeV, the
transition probability becomes unsuppressed.

When the temperature is sufficiently below $T_{\rm c}$, most
transitions are passing the energy barrier close to the
sphaleron and the transition rate $\Gamma$ can be calculated in a
saddle point approximation.

For $T\sim T_{\rm c}$ this approximation breaks down and for $T>
T_{\rm c}$ there is no sphaleron solution at all. Therefore one has to
employ non-perturbative methods to calculate the rate for this
case. It has become popular to call the rate for $T> T_{\rm c}$ the
{\em hot sphaleron rate}.  The idea how to compute it is the
following: The Chern-Simons number is expected to perform a random
walk. Then, for large $t$, thermal average of $\[N_{\rm CS} (t) -
N_{\rm CS} (0) \]^2 $ grows linearly with time,
\begin{eqnarray}
        \left\langle \[N_{\rm CS} (t) - N_{\rm CS} (0) \]^2 \right\rangle
        \to V t \Gamma
        \mlabel{csdiffusion}
\end{eqnarray}
where $V$ is the space volume. The coefficient $\Gamma$ is the
probability for a $N_{\rm CS}$-changing transition per unit time and
unit volume.  The rate for baryon number violation $\Gamma_{\Delta B}$
is proportional to $\Gamma$.  Thus $\Gamma_{\Delta B}$ can be obtained
by computing the real time correlation function \mref{csdiffusion}. At
very high temperature $T\gg T_{\rm c}$ the thermal mass of the Higgs
field becomes large so that it decouples. Then it is sufficient to
evaluate \eq{csdiffusion} in the pure gauge theory.

For the following discussion it is important to identify both the
relevant length scale $R$ of the problem and the relevant size of the
gauge field fluctuations $\Delta\vA$ \cite{khlebnikov}. First of all,
we have to consider fluctuations which are large enough to make a
change $\Delta N_{\rm CS}$ of order unity. To estimate $\Delta N_{\rm
CS}$, recall that $F\tilde{F} = \vE\mal \vB$.  For the electric fields
we have $\vE = - \dot{\vA}$ in $A_0 = 0 $ gauge. Thus $\int dt \vE
\sim \Delta\vA$. Note that the relevant time scale drops out in this
estimate.  Furthermore, we estimate $\vB \sim \Delta\vA/R$ and $\int
d^3 x\sim R^3$. Thus we need
\begin{eqnarray}
\Delta N_{\rm CS} \sim
g^2  \int d^4 x \, 
\vec{E}\cdot\vec{B}
\sim g^2 R^2 (\Delta\vec{A})^2 
\stackrel{!}{\sim} 1
        \mlabel{deltachern}.
\end{eqnarray}
In addition, we have to require that the energy of the relevant fluctuations 
does not exceed the temperature. Otherwise the transition rate
would be Boltzmann suppressed. Proceeding as above we require
\begin{eqnarray}
{\rm energy} \sim \int d^3 x \,
\vec{B}^2
\sim R \, (\Delta\vec{A})^2 
\stackrel{!}{\lsim} T
        \mlabel{energy}.
\end{eqnarray}
Combining Eqs.\ \mref{deltachern} and \mref{energy} we find
\begin{eqnarray}
	R \gsim (g^2 T)^{-1} \qquad 
	\Delta\vec{A} \sim (g R)^{-1}.
\end{eqnarray}
On the other hand, we know that the correlation length of magnetic
fields in a hot non-Abelian plasma is of order $(g ^2 T)^{-1}$. Thus
we can have only $R\lsim (g^2 T)^{-1}$ which finally gives
\begin{eqnarray}
R \sim  (g^2 T)^{-1} \qquad 
\Delta\vec{A} \sim g T
        \mlabel{scale}.
\end{eqnarray}

As I stated above, the relevant time scale $t$, which we need to know in
order to estimate the hot sphaleron rate, has dropped out in these
considerations. It has to be determined from the dynamics and it cannot be
obtained from purely thermodynamic considerations.  Once we know $t$
we can estimate the hot sphaleron rate as
\begin{eqnarray}
	\Gamma \sim t^{-1} R^{-3}
	\mlabel{rate}.
\end{eqnarray}
While \eq{scale} has been well established for a quite
a while, the characteristic time scale has been understood only
recently \cite{letter}. I hope that, by the end of my talk, you will be
convinced that the correct answer is given by \eq{timescale}.

\section{Non-perturbative physics at high temperature}

After the discussion of electroweak baryon number vio\-lation we now
return to our original problem, which is to calculate correlation
functions like \mref{c}.

In the previous section we have seen that the relevant length scale
for electroweak baryon number violation is $(g^2 T)^{-1}$
corresponding to momenta $p\sim g^2 T$.  It is well known that finite
temperature perturbation theory in non-Abelian gauge theories breaks
down for momenta as small as $g^2 T$ \cite{linde80,gross}.  One can see
this by considering the typical size of the transverse gauge field
fluctuations which can be estimated from the equal time correlation
function. For the transverse gauge fields we have
%
%
\begin{eqnarray}
	A_\tr(\vx) \sim g T 
	\mlabel{sizeax}.
\end{eqnarray}
Thus the two terms in the covariant derivative
$\partial_i - i g A_i$ are of the same size which makes perturbation
theory impossible. Note that the thermal fluctuations
are of the same size as the change $\Delta \vA$ which we have estimated
in \eq{scale}.

When I said that perturbation theory does not work for soft momenta,
one may ask: What about the plasmon at rest which has $\vp = 0$ and
the celebrated calculation of it's damping rate \cite{damping}? The
point is that the plasmon oscillation has an amplitude $\sim g^2 T$,
i.e., it is a small oscillation.  What we are interested in is the
time evolution which makes the field change by an amount equal
to it's typical size $\sim g T$.

For the case of equal time correlation functions the non-perturbative
physics associated with the scale $g^2 T$ is relatively well
understood.  It is determined by an effective 3-dimensional pure gauge
theory. This effective theory is the result of integrating out the
field modes with non-zero Matsubara frequency (``dimensional
reduction'') and the 0-component of the gauge fields
\cite{farakos,nieto}.  For sufficiently small coupling this can be
done in perturbation theory. The 3-dimensional theory can be used for
3-dimensional Euclidean lattice simulations which allow for a much
higher precision than 4-dimensional ones \cite{lattice}.  It should be
noted that the use of dimensional reduction is more a matter of
practical convenience rather than a matter of principle. The full
4-dimensional can also be evaluated directly in a 4-dimensional
Euclidean lattice simulation.

The situation is different for unequal time correlation functions.
Dimensional reduction is not possible because we have to consider time
dependent quantities. 4-dimensional Euclidean lattice simulations
are of no use either: Formally, real time correlation functions can be obtained
by an analytic continuation of their imaginary time counterparts.
However, it is not possible to do an analytic continuation of a
function which can be evaluated only numerically.

One simplification of the problem is that the soft field modes behave
classically. This is due to the fact that the momentum scale of
interest is small compared to the temperature. Then the number of
field quanta in one mode with wave vector $\vp$, given by the
Bose-Einstein distribution function
\begin{eqnarray} 
	n(p)=
	\frac{1}{e^{p/T} - 1} \simeq \frac{T}{p} 
	,
\end{eqnarray}
is large.
In this case we are close to the classical
field limit. Thus the dynamics of the soft field modes should be
governed by classical equations of motion. \footnote{It should be 
noted that this argument is somewhat handwaving and it is not easy to 
justify quantitatively \cite{hbar,anharmonic,wb}.}

The first estimate of the hot sphaleron rate, due to Khlebnikov and
Shaposhnikov \cite{khlebnikov}, was based on the assumption that only
the soft fields play a role while the high momentum modes decouple at
leading order. Then the only scale in the problem would be $g^2 T$ and, on
dimensional grounds, the hot sphaleron rate would have to be of the form 
\begin{eqnarray}
	\Gamma_{\rm KS} =\kappa (g^2 T)^4
	\mlabel{ks}
\end{eqnarray}
with a non-perturbative numerical coefficient
$\kappa$. Subsequently attempts were made to compute $\kappa$ by solving the
classical Hamiltonian equations of motion for the gauge fields and
measure $\Gamma$.

However, it turns out that in the case of unequal time correlation
functions is much more complex than in the equal time case.  The main
difficulty is to understand the role of the different momentum scales
which are present in the problem. We will see that both momenta of
order $T$ and momenta of order $gT$ are
relevant to the leading order dynamics of the soft fields.

Therefore one has to integrate out the high momentum modes to obtain
an effective classical theory. 
Schematically, the equations of motion
will be of the form
\begin{eqnarray}
	\frac{\delta S_{\rm eff}[A]}{\delta A_\nu(x)} = 0
        \nn
\end{eqnarray}
where the effective action $S_{\rm eff}[A]$ consists of the tree level
action and the terms which are generated by integrating out the field
modes with $p\sim T$ and $p\sim gT$. Deriving this equation of motion
will be the main subject of my talk.

The reason why the hard modes have to be integrated out to obtain an
effective classical theory for the soft modes is pretty obvious: For
$p\sim T$ the Bose-Einstein distribution is of order 1 so that these
modes certainly do not behave classically. For semi-hard momenta,
however, we have $n(p)\ll 1$.  The main reason why
they have to be integrated out as well is that it does not seem to be 
possible to construct a lattice formulation of the effective theory
for momenta of order $gT$ and $g^2 T$ which allows to take the continuum
limit (cf.\ Ref.\ \cite{bms}). Another reason is that after
integrating out the scale $gT$ we will obtain an effective theory for
the soft field modes which contains only one length scale and one time
scale.  Then it will be trivial to estimate the parametric form of the
hot sphaleron rate.

\section{Integrating out the hard modes}

The hard modes constitute the bulk of degrees of freedom in the hot
plasma. Their physics is that of almost free massless particles moving
on straight lines. Nevertheless, they have a significant influence on
the soft dynamics because they are so numerous.

Integrating out the hard modes means that we have to calculate loop
diagrams with external momenta $\ll T$ and internal momenta of order
$T$. This generates effective propagators and vertices for the field
modes with $p\ll T$.  To leading order, we can restrict ourselves to
one loop diagrams and we can make a high energy or eikonal
approximation of the integrand. The result is nothing but the so
called hard thermal loops \cite{bellac}. The generating functional of the
hard thermal loops has the remarkable property to be gauge invariant.

The main difference between QED and non-Abelian theories is that for
the former the only hard thermal loop is the 2-point function, while
for the latter there are also hard thermal loop $n$-point functions
for all $n$. As we will see later, this has a significant effect on
the soft dynamics, it is in fact qualitatively different in Abelian
and non-Abelian theories.

For the moment, however, we restrict the discussion to the hard
thermal loop 2-point function. This will make clear that the hard
modes have a significant influence on the soft dynamics which was
pointed out by Arnold, Son, and Yaffe \cite{asy}. 

We have to consider the transverse
(or magnetic) components of the gauge fields. The transverse
polarisation function is given by 
\begin{eqnarray}
        \delta \Pi_\tr (P) = \frac12 \mmdebye\[\frac{p_0^2}{p^2}
        + p_0\(1 - \frac{p_0^2}{p^2}\) \intv{}\frac{\! 1}{v\mal P}\]
        \mlabel{deltapi}
\end{eqnarray}
where 
$\mmdebye$ is the leading order Debye mass squared. In a pure
SU($N$) gauge theory, it is given by 
\begin{eqnarray}
	\mmdebye = \frac13 N g^2 T^2. 
\end{eqnarray}
In the hot electroweak theory it receives additional contributions due
to the Higgs and fermion fields. Furthermore, $v^\mu \equiv
(1,\vec{v})$, and the integral $\int d\Omega_{\vv}$ is over the
directions of the unit vector $\vv$, $|\vv| =1$ .

For the naive ``natural'' scale of the problem, i.e., $p_0$, $p\sim g^2 T$
we have
\begin{eqnarray}
	\delta \Pi_\tr (P) \sim g^2 T^2\qquad 
	(p_0\sim g^2 T, \,\,p \sim g^2 T),
\end{eqnarray}
i.e., 
$\delta \Pi_\tr $ is much bigger than the tree level kinetic 
term. Therefore $\delta \Pi_\tr$ must be resummed. The resulting 
transverse propagator
\begin{eqnarray}
        \stern\Delta_{ {\rm t}}(P) =     
          \frac{1}{-P^2 + \delta \Pi_{ {\rm t}}(P)}
        \mlabel{propagator} 
\end{eqnarray}
is smaller than the tree level propagator by a factor $g^2$ when
$p_0\sim g^2 T, \,\,p \sim g^2 T$.  That means that for the naive
``natural'' frequency there are only small perturbative fluctuations
which are of the same amplitude as the plasmon oscillation. It also means
that the estimate \mref{ks} cannot be correct.

In order to obtain large non-perturbative fluctuations we obviously
need to make $\delta \Pi_\tr$ smaller. We know that $\delta \Pi_\tr$ vanishes
\footnote{The longitudinal hard thermal loop polarisation tensor stays
large even for $p_0\to 0$. Thus only the transverse fields can produce
large, non-perturbative fluctuations.}  for $p_0 \to 0$.  Thus we have
to consider small values of $p_0$. In the limit $p_0\ll p$ we have
\begin{eqnarray}
        \delta \Pi_\tr (P) \simeq - \frac{i}{4} \mmdebye \frac{p_0}{p}
        \mlabel{deltapilimit}
\end{eqnarray}
which is purely imaginary. The imaginary part of $\delta \Pi_\tr (P)$
reflects the fact that the dynamics of the soft modes is Landau-damped
by the hard modes. In fact, it is over-damping because the damping
term is much larger than the tree level kinetic term
$p_0^2$. In the same limit the transverse propagator \mref{propagator}
becomes
\begin{eqnarray}
         \stern\Delta_\tr(P) \simeq \frac{1}{p^2} 
        \frac{i\gamma_p}{p_0 + i \gamma_p}
        \mlabel{propagator.5.4}
\end{eqnarray}
with
\begin{eqnarray}
        \gamma_p = \frac{4p^3}{\pi\mmdebye}.
\end{eqnarray}
Now we see how small $p_0$ must be in order to obtain the desired
large fluctuations, namely
\begin{eqnarray}
        p_0 \sim \gamma_p \sim g^4 T,
\end{eqnarray}
corresponding to the time scale 
\begin{eqnarray}
	t\sim \gamma_p^{-1} \sim \(g^4 T\)^{-1}.
\end{eqnarray}
Based on these considerations Arnold, Son and Yaffe estimated the hot
sphaleron rate as
\begin{eqnarray}
        \Gamma_{\rm ASY} = \kappa  g^{10} T^4.
\end{eqnarray}
Subsequently, attempts were made by Huet, Son \cite{huet,son} and
Arnold \cite{arnold} to construct a numerical algorithm to compute the
coefficient $\kappa$ on a lattice.

In this discussion it is assumed that interactions do not change
these order of magnitude estimates. As we will see in the next section,
this is not the case: The field modes with momenta of order $gT$ lead
to interactions of the hard modes which cannot be neglected relative
to the hard thermal loops.

\section{Beyond hard thermal loops}
\mlabel{sec:beyond}

I will now argue that the hard thermal loop approximation for the high
momentum modes ($p\gg g^2 T$) is not sufficient to obtain the correct
effective theory for the soft modes.  

Consider a hard thermal loop and imagine adding a self energy
insertion on an internal line. This gives the diagram in Fig.\
\ref{fig:selfenergy}a in which the hard loop momentum $Q$ is on shell,
$Q^2 = 0$. Since the external momentum $P$ is soft, the momentum $Q+P$
is almost on shell.  The order of magnitude of the self energy
insertion is well known from the calculation of the lifetime of hard
particles \cite{bellac}. From loop momenta $K$ of order $gT$ one gets
a contribution $\sim g^2 T^2$. Compared to the hard thermal loop,
diagram \ref{fig:selfenergy}a also contains an additional propagator
which can be approximated as
\begin{eqnarray}
                \frac{1}{(Q+P)^2} \simeq \frac{1}{2 q}\, \frac{\! 1}{v\mal P}
        \sim \frac{1}{T}\, \frac{\! 1}{v\mal P}
        \mlabel{lee}.
\end{eqnarray}
Thus we can estimate the diagram in Fig.\ \ref{fig:selfenergy}a as
\begin{eqnarray}
        \Pi^{\rm (a)}(P) \sim
         \delta \Pi_\tr(P) \times 
        \frac{g^2 T}{v\mal P}
        \mlabel{pia}
\end{eqnarray}
and, by power counting, this diagram is not suppressed relative to the
hard thermal loop \mref{deltapi} when $P\sim g^2 T$. Note that in this
estimate it did not play any role that we are considering a
non-Abelian theory. The same type of diagram is also present in QED
and the estimate would be exactly the same. However, there is an
essential difference between the Abelian and the non-Abelian case
which will become clear in a moment.

\begin{figure}
 
\vspace*{-2.0cm}
 
\hspace{1cm}
\centerline{
        \hspace{-2cm}
        \epsfysize=16cm\epsffile{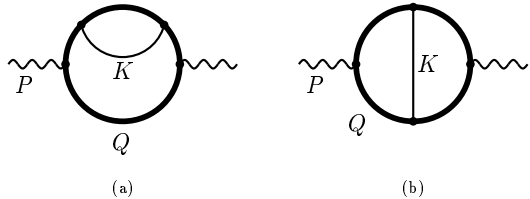}
        }
\vspace*{-10.5cm}
\caption[a]{Two loop contribution to the polarisation tensor for soft
external momentum $P$ ($p_0$, $p \lsim g^2 T$).  The momentum $Q$ is
hard ($q_0$, $q \sim T$). The thick lines denote propagators with
momenta of order $T$. The thin lines are hard thermal loop resummed
gauge field propagators carrying momentum $K$  with $k_0$, $k \sim g T$.
} \mlabel{fig:selfenergy}
\end{figure}

Similarly, one can see that the ladder-type diagram
\ref{fig:selfenergy}b is as large as the hard thermal loop as
well. Again, this estimates are the same in QED and QCD.  However, in
QED the large contributions from the two diagrams turn out to be the
same with opposite signs and cancel.

I will now argue that this cancellation does not occur in a
non-Abelian theory.  Let me first give a formal explanation in terms
of diagrams. At the end of Sect.\ \ref{sec:logarithmic}, I will come
back to this point and I will give a more intuitive argument.

Here is the diagrammatic argument: We have seen that the two diagrams
both contain a hard and a semi-hard loop momentum. Imagine now that we
calculate them in two steps: First do the integral over the hard
momentum with $K$ kept fixed. This can be interpreted as follows: The
first step generates an effective 4-point vertex for momenta of
order $gT$ and $g^2 T$.  The result of such a calculation is well
known: At leading order this is nothing but the hard thermal loop four
point function. And we also know that there is no such vertex in QED,
even though individual diagrams would give such a contribution. 
Summing over the permutations of external lines the large contributions
cancel. In a non-Abelian theory each of these diagrams comes with a
different colour factor and the sum over permutations gives a non-vanishing
result. In the
second step we integrate over momenta of order $gT$ which can be done
using the well known expression for the hard thermal loop 4-point
function.

\section{Integrating out the semi-hard modes}
\mlabel{sec:semihard}

We have seen in the previous section that, in order to obtain an
effective theory for the soft momentum fields, we cannot restrict
ourselves to the hard thermal loops. Not only the hard modes but also
the semi-hard ones affect the leading order dynamics of the soft
fields.  At first sight this appears like an additional
complication. Somewhat surprisingly, things become much simpler
instead.

We have also seen that it is convenient
to proceed in two steps: In the first step we integrate out momenta
of order $T$ which yields the hard thermal loop effective theory. In the
second step this effective theory is used to integrate out the scale
$gT$. In this way 
the calculation 
will be simplified  because
we have a partial cancellation, corresponding to the QED-type
contributions, already ``built in''.  

\subsection{One loop diagrams}
\mlabel{sec:one.loop}

\begin{figure}[t]
 
\vspace*{-2.0cm}
 
\hspace{1cm}
\centerline{\epsfysize=16cm\epsffile{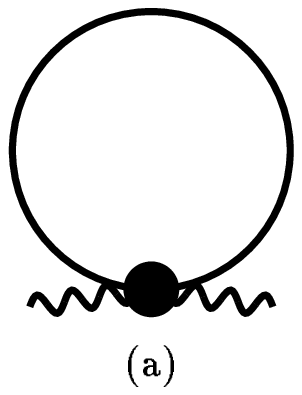}
        \hspace{-8cm}\epsfysize=16cm\epsffile{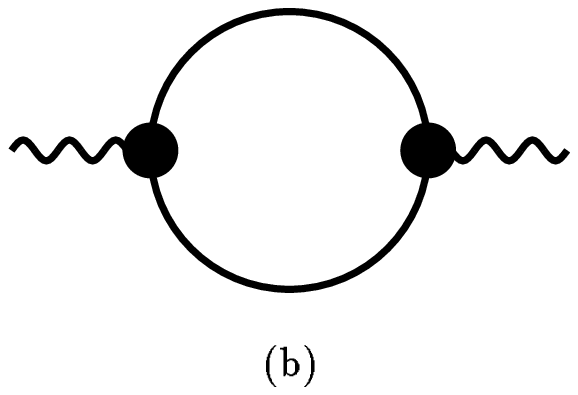}
        }

\vspace*{-11cm}
 
\caption[a]{One loop  contributions to the polarisation tensor 
  in the ``hard thermal loop effective theory''. The heavy dots are
  hard thermal loop vertices. Otherwise the notation is the same as in
  Fig.\ \ref{fig:selfenergy}. }

\mlabel{fig:htlvertex}
\end{figure}

We start by considering the one loop diagram in the hard thermal
thermal loop effective theory depicted in Fig.\ \ref{fig:htlvertex}a.
It corresponds to the sum of the diagrams of the original theory in
Fig.\ \ref{fig:selfenergy}. We neglect the external soft momentum
relative to the semi-hard one whenever it's possible. In this way we
obtain an expression in which the loop integration is logarithmically
infrared divergent. This should not bother us since we do not want to
compute the diagram exactly. All we want to do is to integrate out
momenta of order $g T$. Therefore we need a scale $\mu$ which
separates semi-hard momenta from soft ones,
\begin{eqnarray}
        g^2 T\ll \mu \ll gT.
\end{eqnarray}
This scale
appears as a lower limit in our integral so that it is no longer
divergent. The integral is difficult evaluate. What it not
so difficult, is to compute the terms which are singular for $\mu \to 0$.
One finds \cite{eff}
\begin{eqnarray}
         \Pi_{\mu\nu}^{(4)} (P) &=& -\frac{i}{4\pi}
         \mmdebye N g^2 T \log\(\frac{gT}{\mu}\) \nn \\ &&
        \times p_0 \intv{} \frac{v_\mu v_\nu
        }{(v\mal P)^2}\,\, 
      \mlabel{pi4}.
\end{eqnarray}
This result is gauge fixing independent in a general covariant gauge.

Obviously, \eq{pia} is not suppressed relative to the hard thermal loop
selfenergy \mref{deltapi}: The prefactor $g^2 T$ is compensated by
one additional factor $v\mal P \sim g^2 T$ in the
denominator. \eq{pia} is even larger than \mref{deltapi} by a factor of
$\log(gT/\mu)$.

Note that \mref{pi4} is not transverse, which would be necessary for
the effective theory, which we are going to derive, to be gauge
invariant. There is, however, another one loop diagram in the hard
thermal loop effective theory which is of the same size as \eq{pia}
and which is depicted in Fig.\ \ref{fig:htlvertex}b.  With the same
approximations as above one obtains \cite{eff}
\begin{eqnarray}
        \Pi_{\mu\nu}^{(3)} (P) &=& 
        \frac{i}{\pi^2} \mdebye^2 N g^2 T \log\(\frac{gT}{\mu} \)
        \nn\\ && {}
        \hspace{-2cm}
        \times p_0
        \intv{1}\frac{v_{1\mu}}{v_1\mal P} \intv{2} \frac{v_{2\nu}}{v_2\mal P}
        \frac{(\vv_1\mal\vv_2)^2}{\sqrt{1 - (\vv_1\mal\vv_2)^2}}
        \mlabel{pi3}.
\end{eqnarray}
One easily verifies that the sum of Eqs.\ \mref{pi4} and \mref{pi3}
is transverse,
\begin{eqnarray}
	P^\mu\( \Pi_{\mu\nu}^{(4)} (P) + \Pi_{\mu\nu}^{(3)} (P)\) = 0.
\end{eqnarray}
\subsection{Higher loops}

There are higher loop diagrams in the hard thermal loop effective
theory which are not suppressed relative hard thermal loops either.
Obviously these terms need to be resummed.  As I already said, the use
of the hard thermal loop effective theory has simplified the
calculation of \mref{pi4} and \mref{pi3} quite a bit. Nevertheless,
higher loop diagrams are still difficult to calculate because the
number of terms in the hard thermal loop $n$-point functions grows
rapidly with increasing $n$.

It is much more efficient to use another formulation of the hard
thermal loop effective theory, which is in terms of kinetic
equations. The degrees of freedom in this description are the gauge
fields $A_\mu(x)$ and, in addition, the fields $W(x,v)$ which
transform under the adjoint representation of the gauge group. The
fields $W(x,v)$ depend both on the space-time coordinates and on the
unit vector $\vv$. They describe the deviation of the distribution of
hard particles with 3-velocity $\vv$ from
thermal equilibrium. 

The equations of motion for these fields are the non-Abelian generalisation
of the Vlasov equations for a QED plasma \cite{blaizot93},
\begin{eqnarray}
        \mlabel{maxwell}
        [D_\mu, F^{\mu\nu}(x)]=  \mmdebye \int\frac{d\Omega_\vec{v}}{4\pi}
        v^\nu  W(x,v),
\end{eqnarray}
\begin{eqnarray}
        [v \cdot D, W(x,v)] = \vec{v}\cdot\vec{E} (x)
        \mlabel{vlasov},
\end{eqnarray}
where $\vec{E}$ is the non-Abelian electric field, and the 4-vector $v$
is defined as in \eq{deltapi}.  The rhs of \eq{maxwell} is the current
due to the hard particles.  The conserved Hamiltonian corresponding to
Eqs.\ \mref{maxwell} and \mref{vlasov} is \cite{nair,blaizot94}
\begin{eqnarray}
         H&=&\int d^3 x \trace \Bigg\{
        \vec{E}(x)\cdot\vec{E}(x) +\vec{B}(x)\cdot\vec{B}(x) \nn\\ &&
        \hspace{1cm}
        {}+
        \mmdebye\int\frac{d\Omega_\vec{v}}{4\pi}
        W(x,v) W(x,v)\Bigg\}
        \mlabel{hamiltonian}.
\end{eqnarray}
One advantage of this formulations that this
theory is local. What is even more important, however, is that the
algebraic complexity is significantly reduced.

What does it mean to integrate out the semi-hard field modes in this
formulation?  First we have to split the fields into two pieces.  
The fields $A$, $\vec{E}$ and $W$ are decomposed into soft
and a semi-hard  modes \footnote{For notational simplicity I do not introduce
new symbols for the soft modes. From now on $A$, $\vec{E}$ and $W$
will always refer to the soft fields only.},
\begin{eqnarray}
   A &\to& A + a\nn \\
   \vec{E} &\to& \vec{E} + \vec{e}\nn \\
   W &\to& W + w 
   \mlabel{separation}.
\end{eqnarray}
The soft modes $A$, $\vec{E}$ and $W$ contain
the spatial Fourier components with $p<\mu$ while the semi-hard modes
$a$, $\ve$ and $w$ consist of those with $k>\mu$. 

The interpretation of $W$ and $w$ is the following: Both describe
deviation of the distribution of the hard particles from thermal
equilibrium. $W$ ($w$) is the slowly (rapidly) varying piece of
this distribution varying on length scale greater (less) than $1/\mu$.

After this split we have two sets of  equations of
motion, one for the soft fields and one for the semi-hard fields. 
Due to the non-linear terms these sets are coupled. 

The equations for the soft fields can be written as
\begin{eqnarray}
        [D_\mu, F^{\mu\nu}(x)] &=&  \mmdebye \int\frac{d\Omega_\vec{v}}{4\pi}
        v^\nu  W(x,v)
        \mlabel{maxwellsoft},
\end{eqnarray}
\begin{eqnarray}
 	\[ v \cdot D, W(x,v) \] &=& \vec{v}\cdot\vec{E} (x)
        + \xi(x,v) ,
        \mlabel{vlasovsoft}
\end{eqnarray}
where \footnote{The structure constants $f^{abc}$ are defined such
that $[T^a,T^b] = i f^{abc} T^c$.}
\begin{eqnarray}
        \xi^a (x,v) = -g f^{abc} \( v\cdot a^b (x)
        w^c(x,v) \)_{\rm soft} .
        \mlabel{xi}
\end{eqnarray}
The subscript ``soft'' indicates that only spatial Fourier
components with $p<\mu$ are included.  The field strength tensor and
the covariant derivatives in Eqs.\ \mref{maxwellsoft} and
\mref{vlasovsoft}  contain only the soft gauge fields.  Note that
terms which couple soft and semi-hard fields have
been neglected in \eq{maxwellsoft}. 

Since the semi-hard modes can be treated perturbatively,
one can neglect their self-interaction. One can also neglect
interactions in the first Vlasov equation, so that we have
\begin{eqnarray}
        \partial_\mu f^{\mu\nu}(x) &=&
        \mmdebye \int\frac{d\Omega_\vec{v}}{4\pi}
        v^\nu  w(x,v)
        \mlabel{maxwellhard},
\end{eqnarray}
\begin{eqnarray}
        v\mal \partial w(x,v) &=&
         \vv\mal\ve   (x) +   h (x,v),
        \mlabel{vlasovhard}
\end{eqnarray}
where $f^{\mu\nu} = \partial^\mu a^\nu - \partial^\nu a^\mu$.
The only interaction which affects the time evolution of the
semi-hard fields at leading order appears in \eq{vlasovhard}, 
it is due to the term
\begin{eqnarray}
         h^a (x,v) &=&
        -g f^{abc}\Big[ v\mal A^b(x) w^c(x,v) \nn\\ &&
                  \hspace{2cm} {} +v\mal a^b(x) W^c(x,v)\Big].
\end{eqnarray}

Integrating out the scale $gT$ means that we solve the classical
equations of motion \mref{maxwellhard} and \mref{vlasovhard} for the
semi-hard fields on the soft background. Then we have to perform the
thermal average over their initial conditions using the Hamiltonian
\mref{hamiltonian}. In this way we eliminate the semi-hard fields and
we end up with equations of motion for the soft fields only. This can
be done in perturbation theory, where one expands in powers of
$h(x,v)$.  We obtain an expansion
\begin{eqnarray}
        a(x) &=& a_0(x) + a_1(x) + a_2(x)  + \cdots\nn\\
        w(x,v) &=& w_0(x,v) + w_1(x,v) + w_2(x,v) + \cdots
        \mlabel{solution}
\end{eqnarray}
in which the $n$-th order term contains $n$ powers of the soft fields
$A$, $W$. The terms $a_0$, $ w_0$ are independent of the soft
background and they only depend on the initial conditions for the
semi-hard fields.  Note that the solution \mref{solution} is still
formal since it contains the yet unknown non-perturbative soft fields.
 
Inserting \mref{solution} into \eq{xi} we obtain a series
\begin{eqnarray}
        \xi(x,v) = \xi_0(x,v) + \xi_1(x,v) + \xi_2(x,v) + \cdots
        \mlabel{xiexpansion}.
\end{eqnarray}
Each term in \mref{xiexpansion} is bilinear in free fields $a_0$ and
$w_0$. The term $\xi_n$ is of $n$-th order in the fields $A$ and $W$.
This expression has to be plugged into the second Vlasov equation
\mref{vlasovsoft} for the soft fields, which then becomes
\begin{eqnarray}
        [v \cdot D, W(x,v)] &=& \vec{v}\cdot\vec{E} (x)\nn\\ && 
        \hspace{-2cm}
        {}+ \xi_0(x,v) + \xi_1(x,v) + \xi_2(x,v) + \cdots
        \mlabel{vlasovsoft.2}.
\end{eqnarray}

As it stands, \eq{vlasovsoft.2} is not very useful yet. 
On the rhs we have an infinite series of terms, each depending
on the yet unknown soft fields and on the initial conditions
for the semi-hard fields. In general, the average over these
initial conditions can be performed only after the equations
for the soft fields have been solved. 

Fortunately, we can simplify \eq{vlasovsoft.2} significantly since we
are only interested in the leading order dynamics of the soft field
modes. Recall that each term $\xi_n$ is bilinear in the fields $a_0$
and $w_0$. Furthermore, $a_0$ and $w_0$ are linear in the initial
values for their time evolution. After solving the equations of motion
for the soft fields \mref{maxwellsoft} and \mref{vlasovsoft.2}, one has
to average over initial conditions.  Due to the scale separation the
average of the bilinear terms of semi-hard fields can be approximated
by disconnected parts.  But this just means that we can perform the
average over initial conditions for the semi-hard fields already in
the equation of motion \mref{vlasovsoft.2}! The only exception is the
term $\xi_0$ because the thermal average of this term vanishes due to
colour conservation: Thermal averages like $\langle a_0^a a_0^b\rangle$
are proportional to $\delta^{a b}$. Inserted into \eq{xi} this gives
zero due to the antisymmetry of the structure constants $f^{abc}$.
The highest correlation function of $\xi_0$ we have to take into
account is the two point function $\langle \xi_0(x_1,v_1)
\xi_0(x_2,v_2)\rangle$. Put differently, the term $\xi_0$ acts like a
Gaussian random force. It is random because it is independent of the
soft fields.

Since the expectation value of $\xi_0$ vanishes, it is not sufficient
to keep only this term on the rhs of \eq{vlasovsoft.2}. We also
have to take into account $\xi_1$ which we
can approximate by
\begin{eqnarray}
        \xi_1(x,v) \simeq \lav \xi_1(x,v) \rav
        \mlabel{simplify.xi1}
\end{eqnarray}
where $\lav \cdots \rav$ denotes the average over initial conditions
for the semi-hard fields. The higher order terms $\xi_n$ in
\eq{vlasovsoft.2} can be neglected.  Then we have
\begin{eqnarray}
        [v \cdot D, W(x,v)] &\simeq& \vec{v}\cdot\vec{E} (x)\nn\\ && 
        \hspace{-2cm}
        {}+ \xi_0(x,v) + \lav\xi_1(x,v)\rav 
        \mlabel{vlasovsoft.3}
\end{eqnarray}
in which the correlation functions of $\xi_0$ can be treated as Gaussian.

\section{The logarithmic approximation} 
\mlabel{sec:logarithmic}

The next step towards the effective equations of motion for the soft
fields is to evaluate the terms in the second line of
\eq{vlasovsoft.3}.  As in the diagram calculation in Sec.\
\ref{sec:one.loop} one encounters terms which are logarithmically
sensitive to the separation scale $\mu$.  Only these logarithmic terms
will be kept. The reason why this  is sufficient will
become clear in a moment.

We have seen that all we need to know about
the noise term $\xi_0$ is its 2-point function. The
leading logarithmic result reads 
\begin{eqnarray}
        \Big\langle 
    \xi_0^{a}(x_1,v_1)
    \xi_0^{b}(x_{2},v_{2}) 
    \Big\rangle 
    &=&
    -2 N \frac{g^2 T^2}{\mmdebye}
        \log\(\frac{g T}{\mu}\)
         I(v_1,v_2)\nn\\ && \times
        \delta^{ab} \delta^{(4)}(x_1 - x_2),
        \mlabel{xi0correlator4}
\end{eqnarray}
with
\begin{eqnarray}
  \mlabel{k}
  I(v,v_1) \equiv -\deltav(\vv - \vv_1) 
        + \frac{1}{\pi^2}
        \frac{(\vv\cdot\vv_1)^2}{\sqrt{1 - (\vv\cdot\vv_1)^2}} ,
\end{eqnarray}
where $\deltav$ is the delta function on the two dimensional unit
sphere: 
\begin{eqnarray}
        \int d\Omega_{\vv_1} f(\vv_1) \deltav(\vv - \vv_1) = f(\vv) .
\end{eqnarray}
With the same approximations the result for the last term in
\eq{vlasovsoft.3} is
\begin{eqnarray}
        \lav\xi_1(x,v)\rav &=&
         N g^2 T  \log\(\frac{g T}{\mu}\)\nn\\&&\times
        \intv{1}   I(v,v_1)      
        W(x,v_1). 
\end{eqnarray}
Inserting this  into \eq{vlasovsoft.3}
we obtain 
\begin{eqnarray}
        [v \cdot D, W(x,v)] &=& \vec{v}\cdot\vec{E} (x)+ \xi_0(x,v)
        \nn\\ &&
        \hspace{-2cm} {}  + N g^2 T  \log\(\frac{g T}{\mu}\)
 \intv{1}   I(v,v_1)      
 W(x,v_1) . 
        \mlabel{boltzmann}
\end{eqnarray}

I will now argue that, for the leading order dynamics of the soft
fields, the lhs of \eq{boltzmann} can be neglected.  The only spatial
momentum scales which are left in the problem are $\mu$ and $g^2
T$. The field modes we are ultimately interested in, are the ones
which have only momenta of order $g^2T$.  The cutoff dependence on the
rhs must drop out after solving the equations of motion for the fields
with spatial momenta smaller than $\mu$.  Thus, after the
$\mu$-dependence has cancelled, the logarithm must turn into
$\log(gT/(g^2 T)) = \log(1/g)$.

We will now simplify \eq{boltzmann} by neglecting terms which are
suppressed by inverse powers of $\log(1/g)$. We introduce moments of
$W(x,v)$ and $\xi_0(x,v)$,
\begin{eqnarray}
        W^{i_1\cdots i_n} (x) &\equiv& \intv{} v^{i_1} \cdots  v^{i_n} W(x,v)
        ,\nn\\
        \xi_0^{i_1\cdots i_n} (x) &\equiv& \intv{} v^{i_1} \cdots  v^{i_n} 
        \xi_0(x,v) .
\end{eqnarray} 
Taking the first
moment of Eq.~\mref{boltzmann} gives
\begin{eqnarray}
        [D_0,W^i(x)] - [D^j,W^{ij}(x)] &=& \frac13 E^i(x) \nn\\
        && \hspace{-3.5cm} {}+ \xi_0^i(x) 
        -  \frac{N g^2 T}{4\pi}  \log(1/g) W^i(x)
        \mlabel{wi.2}.
\end{eqnarray}
Assuming that (cf.\ Ref.\ \cite{eff}) $W^{ij}$ is of the same size as
$W^i(x)$ and $D_0\lsi g^2 T$, the lhs of Eq.~\mref{wi.2} is
logarithmically suppressed relative to the term $\propto W^i(x)$ on the rhs and
can be neglected. Then we can trivially solve \mref{wi.2} for $W^i(x)$,
\begin{eqnarray}
   W^i(x) = \frac{4\pi}{N g^2 T  \log(1/g)} \(\frac13 E^i(x) + \xi_0^i(x) \).
\end{eqnarray}
Inserting the current $j^i(x) = \mmdebye W^i(x)$ into the rhs of
\eq{maxwellsoft} gives
\begin{eqnarray}
        [D_\mu,F^{\mu i}(x)] = 
        \gamma E^i(x) + \zeta^i(x)
        \mlabel{langevin1}
\end{eqnarray}
where we have introduced 
\begin{eqnarray}
        \zeta^i(x) \equiv  \frac{4\pi\mmdebye}{Ng^2T\log(1/g)}
         \xi_0^i(x)
\end{eqnarray}
and 
\begin{eqnarray}
        \gamma = \frac{4\pi\mmdebye}{3 Ng^2 T\log(1/g)}.
\end{eqnarray}
The term $\zeta$ is a Gaussian white noise.
Its correlator is easily obtained from \eq{xi0correlator4},
\begin{eqnarray}
        \Big\langle 
    \zeta^{ia}(x_1)
    \zeta^{jb}(x_{2}) 
    \Big\rangle 
        =
    2 T \gamma
        \delta^{ij } \delta^{ab}  \delta^{(4)}(x_1 - x_2)
        \mlabel{langevin2}.
\end{eqnarray}

\eq{langevin1} is the main result of our calculation.  It is
remarkable that, at the order we are considering, the effect of the
high momentum modes can simply be described by a local damping term,
together with a Gaussian white noise which keeps the soft modes in
thermal equilibrium.  This surprising result is somewhat
counter-intuitive since Landau damping is generally known as a
non-local effect. Qualitatively, it can be understood as follows:
Recall that the hard particles have a mean free path of order $(g^2
T\log(1/g))^{-1}$ \cite{bellac}. To understand the physical picture
behind the logarithmic approximation, we have to imagine that $\log
(1/g)$ is a large number.  Then the mean free path of the hard
particles is small compared to the length scale $(g^2 T)^{-1}$ on
which the soft non-perturbative dynamics occurs. That is, the soft
field modes are Landau damped only over a small length scale.

There is an essential difference compared to Abelian gauge theories.
There the mean free path is of order $(g^2 T\log(1/g))^{-1}$ as well.
However, the mean free path is determined by the total cross section
of the hard particles which is dominated by small angle scattering.
Thus, in a typical scattering event, the momentum of a charged
particle in a QED plasma is hardly affected. The Landau damping can
continue as a coherent process after such a scattering occurs. In a
non-Abelian theory even a scattering under arbitrarily small angle can
change the colour charge of a hard particle. It is precisely this colour
exchange which leads to a loss of coherence in the process of Landau
damping.

\section{The hot sphaleron rate}

With \eq{langevin1} we are now able to estimate the characteristic
time scale $t$ of non-perturbative gauge field fluctuations and thus
of the rate for hot electroweak baryon number violation. Neglecting
the term $[D_0,F^{0 i}(x)]$ (see below) the lhs can be estimated as
\begin{eqnarray}
	[D_j,F^{j i}(x)] \sim R^{-2} \Delta \vA\sim g^4 T^2 \Delta \vA
	\mlabel{lhs}.
\end{eqnarray}
On the rhs we have in $A_0=0$ gauge
\begin{eqnarray}
	\gamma \vE \sim \frac{T}{\log(1/g)} \frac{\Delta \vA}{t}
	\mlabel{rhs}.
\end{eqnarray}
Putting \mref{lhs} and \mref{rhs} together we find 
\begin{eqnarray}
  t\sim \(g^4 \log(1/g)T \)^{-1}
	\mlabel{timescale}.
\end{eqnarray}
Therefore the hot sphaleron rate, at leading order, has the form 
\begin{eqnarray}
  \Gamma = \kappa g^{10}\log(1/g)T^{4}
	\mlabel{rate.2}
\end{eqnarray}
where $\kappa$ is a non-perturbative coefficient which does not depend
on the gauge coupling.

\section{An effective theory for the soft modes}
\mlabel{sec:effective}

I will now discuss how the non-perturbative coefficient $\kappa$ in
\eq{rate.2}, and more generally, correlation functions like \mref{c}
can be calculated on a lattice for which it is convenient to work
in the $A_0 = 0$ gauge. 

The time scale for non-perturbative dynamics is much
larger than the corresponding length scale. Therefore time
derivatives on the lhs of \eq{langevin1} are negligible and
the non-perturbative dynamics of the soft gauge fields
at leading order is correctly described by the equation
\begin{eqnarray}
        [D_j,F^{j i}(x)] = 
        -\gamma \dot{A}^i(x) + \zeta^i(x)  
        \mlabel{langevintag},
\end{eqnarray}
which should be easy to implement in a lattice calculation.

After solving \eq{langevintag}, the result has to be plugged into the
operator ${\cal O}(t)$ of interest.  The corresponding correlation
function is then given by the average over the random force $\zeta$
using \eq{langevin2}.

\section{Summary and discussion}
\mlabel{sec:summary}

We have obtained an effective theory for the non-perturbative dynamics
of the soft field modes by integrating out the hard ($p\sim T$) and
semi-hard modes ($p\sim gT$) in perturbation theory.  This effective
theory is described by the Langevin equation \mref{langevintag}.

Furthermore, we have determined the parametric form of the hot
electroweak baryon number violation rate at leading order. It contains a
non-perturbative numerical coefficient which can be evaluated using
\eq{langevintag}.

One would expect corrections to \eq{rate.2} to be suppressed by a factor
$(\log(1/g))^{-1}$, which, in the electroweak theory, is not small
\footnote{We have $g\simeq 0.66$ for
$T\sim 100$GeV which gives $\log (1/g)\simeq 0.4$.}. It would therefore
be interesting, both from the theoretical and from the practical point of view,
to see how sub-leading terms can be computed.

\acknowledgments

This work was supported in part by the TMR network
``Finite temperature phase transitions in particle physics'', EU
contract no. ERBFMRXCT97-0122.

\vspace*{1cm}
\noindent {\bf Note added:} While this paper was being completed, I
received a preprint by Arnold, Son and Yaffe \cite{asy98}, in which
they derive Eq.\ \mref{langevintag} using the concept of colour
conductivity. They also show that the effective theory described by
this equation is insensitive to the ultraviolet which means that its
lattice implementation does not require any renormalization.

\end{document}